\documentclass[twocolumn,showpacs,preprintnumbers,amsmath,amssymb,prl]{revtex4}

\usepackage{graphicx}
\usepackage{dcolumn}
\usepackage{epsf}

\begin{document}

\title{Size effects in the nonlinear resistance and flux creep in a virtual
Berezinskii-Kosterlitz-Thouless state of superconducting films.}

\author{A. Gurevich$^{1}$ and V.M. Vinokur$^{2}$.}
\affiliation{$^{1}$National High Magnetic Field Laboratory, Florida
State University, Tallahassee, Florida, 32310 \\
$^{2}$Materials Science Division, Argonne National Laboratory,
Argonne, Illinois, 60439}

\date{\today}
\begin{abstract}

We show that the size effects radically affect the electric
field-current ($E-I$) relation of superconducting films.  We
calculate $E(J)$ due to thermally-activated hopping of single
vortices driven by current $I$ across the film in a magnetic field
$H$, taking into account interaction of free vortices with their
antivortex images and peaks in the Meissner currents at the film
edges. Unbinding of virtual vortex-antivortex pairs not only mimics
the transport uniform BKT behavior, it can dominate the observed
$E(J)$ and result in the field-dependent ohmic resistance at small
$I$. We show that $E(I)$ can be tuned by changing the film geometry
and propose experimental tests of this theory.

\end{abstract}

\maketitle

The Berezinskii-Kosterletz-Thouless (BKT) transition is a
2D universal phase transition due to unbinding of logarithmically
interacting topological excitations~\cite{bkt}. The concept of the BKT
transition first introduced in the context of vortices in
$XY$-magnets has been extended to other topological excitations like
vortex-antivortex pairs in superfluid films, superconducting films,
Josephson-junction arrays \cite{hn,minhag}, dislocation pairs in the
theory of 2D melting or ultracold atomic gases in optical lattices
\cite{opt}. The superconducting films and Josephson arrays have
become the main experimental testbeds to study the BKT transition by
dc transport measurements. In this case the ohmic electric field-current
characteristics $R=RI$ above the transition $T>T_{BKT}$ turns into
the power-law $E\propto I^{1+\alpha}$ at $T<T_{BKT}$ with a jump to
$\alpha=2-5$ followed by the growth of $\alpha$ as the temperature
$T$ decreases~\cite{minhag}.

While the interaction of dislocations and vortices in $XY$-magnets
of superfluid films is indeed logarithmic, the interaction of
vortices in superconducting films is only logarithmic over distances
shorter than the Pearl screening length $\Lambda = \lambda^2/d$
where $d$ is the film thickness and $\lambda$ is the London
penetration depth~\cite{pearl}.  The size effects can change the BKT
transport behavior at $T<T_{BKT}$ since the result, $E\propto
I^{1+\alpha}$, holds only at sufficiently high currents, $I >
I_1\sim c\epsilon/\phi_0$, for which the critical size of a
dissociating vortex-antivortex pair, $\ell_c=2cw\epsilon/\phi_0I$,
is smaller than the film width, $w$, where $\alpha=2\epsilon/T$,
$\epsilon=\phi_0^2/16\pi^2\Lambda$ is the vortex energy scale,
$\phi_0$ is the flux quantum, and $c$ is the speed of light. For $I<I_1$, the $E-I$ characteristic becomes
ohmic~\cite{hn,minhag,edge,fin,gia}. Yet several crucial features of
the electrodynamics of superconducting films have not been
incorporated into the BKT theory. First, the sheet current density
$J(x)$, which drives vortices across the film can be highly
nonuniform. For a current-carrying thin film strip of width
$w>\Lambda$ in a perpendicular magnetic field $H$, we have
\cite{eb}:
    \begin{figure}                  
    \epsfxsize= 0.45\hsize
    \centerline{
    \vbox{
    \epsffile{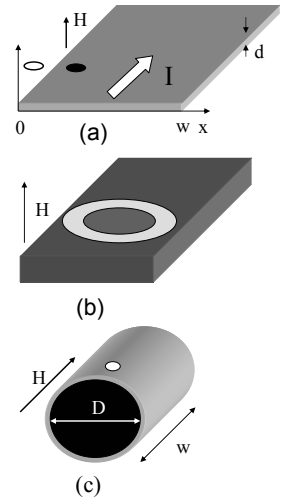}
    }}
    \caption{A thin film in a perpendicular field $H$. The black dot
    shows a vortex moving across the film, and the empty circle shows the antivortex image (a). Geometries for probing
    the resistive state by relaxation measurements: a thin ring in a perpendicular field (b), and a thin tube on a
    cylindrical substrate in a parallel field (c). The white dot shows the vortex driven along the tube by the azimuthal Meissner currents. }
    \label{Fig.1}
    \end{figure}
    \begin{equation}
    J(x)=[I+(w-2x)Hc/4]/\pi\sqrt{x(w-x)},
    \label{meiss}
    \end{equation}
where the geometry is shown in Fig. 1a. This distribution of $J(x)$
ensures no spontaneous vortices generated by small $I$ and $H$ in
the film (the singularities at the film edges are cut off at the
distances $\sim \mbox{max}(d,\lambda)$). The second feature results
from the Bean-Livingston surface barrier: a vortex penetrating a
film interacts with a fictitious antivortex image, which provides
zero normal currents at the edges. Thus, thermally-activated
penetration of single vortices is governed by the BKT-type unbinding
of a virtual vortex - antivortex pair \cite{edge}. For $w<\Lambda$,
the interaction energy $U({\bf r_1}, {\bf r_2})$ between two
vortices is logarithmic only for small separation, $|{\bf r_1}-{\bf
r_2}|<w$, otherwise $U({\bf r_1}, {\bf r_2})$ decays exponentially
over the length $w/\pi$ along the film because of cancelation of the
vortex currents by an infinite chain of vortex-antivortex images
\cite{film}. This makes rare thermally-activated hops of vortices
across the strip uncorrelated at low $T$ and $I$.

In this Letter we show that fluxon hopping mediated by the unbinding
of a vortex from its edge antivortex images mimics the uniform BKT
resistive state and results in a strongly size-dependent $E(I)$,
which can exceed $E_2(I)$ caused by the uniform pair dissociation
\cite{hn} both for $w<\Lambda$ and $w>\Lambda$. This is due to the
fact that the energy activation barrier for the single vortex
penetration is roughly half of the barrier required to create a
vortex-antivortex pair in the film. The account of these features is
important for the interpretation of deviations from the BKT scenario
and critical currents observed on $E-I$ curves of ultrathin films
\cite{lobb,jrt,martin,tafuri}. Since it is the thin film strip
geometry, which is mostly used in dc transport measurements, we also
discuss other geometries in which the genuine BKT pair dissociation
could be revealed.

We calculate $E(I)$ due to vortex hopping across a thin film
described by the Langevin equation $\eta\dot{x}+U^\prime(x)=\zeta$
where the dot and the prime denote differentiations over time t and
coordinate x, respectively, $\eta$ is the viscous drag coefficient,
$\zeta(t)$ describes thermal noise and the local energy
$U(x)=U_0-U_m$ comprises the position-dependent vortex self-energy
$U_0(x)$ and the work of the Meissner current,
$U_m=(\phi_0/c)\int_0^xJ(u)du$ to move the vortex by the distance
$x$ from the film edge. Here $J(x)$ is described by the integral
Maxwell-London equation \cite{pearl,eb}
    \begin{equation}
    \int_0^w\frac{J(u)du}{u-x}+4\pi\Lambda\partial_xJ=-cH
    \label{inteq}
    \end{equation}
supplemented by the condition $I=\int_0^wJ(x)dx$. If $w\gg\Lambda$,
Eq. (\ref{inteq}) yields Eq. (\ref{meiss}), but for $w\ll\Lambda$,
the integral term is negligible, and $J(x)\simeq
I/w+cH(w-2x)/8\pi\Lambda$.

The self-energy $U_0(x)=-\int_0^x F(u)du$ is the work
required to create a vortex at the edge where $U_0(0)=0$ and move it
by the distance $x$. Here
$F(x)=f(2x)+\sum_{n=1}^\infty[f(2wn+2x)-f(2wn-2x)]$ is the
force between the vortex in the film and an infinite chain of vortex
and antivortex images outside the film, $f(x)=\phi_0J_y(x)/c$, and
$J_y(x)$ is the y-component of the sheet current density of the Pearl vortex 
in an infinite film. Using  $J_y(k)=-ic\phi_0k_x/2\pi k(1+2\Lambda k)$,
$k^2=k_x^2+k_y^2$ \cite{pearl} and integrating over $k_y$ in the Fourier space, we obtain:
\begin{equation}
U_0=\frac{\phi_0^2}{\pi^2}\sum_{n=1}^N\frac{\sin^2 (\pi n
x/w)}{\sqrt{(2\pi n \Lambda)^2-w^2}}\tan^{-1}\left[\frac{2\pi n
\Lambda-w} {2\pi n \Lambda+w}\right]^{1/2} \label{uoo}
\end{equation}
Here $N\simeq we^{-C}/2\pi\xi$ and $C=0.577$ provide the vortex core
cutoff. For narrow films $w\ll 2\pi\Lambda$, the summation in Eq.
(\ref{uoo}) reproduces the known result  \cite{film,vgk}:
    \begin{equation}        
    U_0(x)=\epsilon\ln[(w/\pi\xi)\sin(\pi
    x/w)],
    \label{u0}
    \end{equation}
where $\epsilon=\phi_0^2/16\pi^2\Lambda$.  Here $U_0$ results from
the kinetic energy of unscreened vortex supercurrents cut off at the
distance $\sim\xi$ from the edges where the London theory breaks
down. For wide films $w>2\pi\Lambda$, $U_0(x)$ increases from zero at
$x=0$ to $U_a\simeq\epsilon \ln (\Lambda/\xi)$ over the length
$x\sim\Lambda$. The magnetic part of the energy barrier $U_m(x)$ for
$w\gg \Lambda$ and $w\ll \Lambda$, is given by
    \begin{eqnarray}
    U_m=\frac{2\phi_0I}{\pi c}\sin^{-1}\sqrt{x/w}+\frac{\phi_0H}{2\pi}\sqrt{x(w-x)}, \quad w\gg \Lambda
    \label{umb} \\
    U_m=\phi_0Ix/cw+\phi_0Hx(w-x)/8\pi\Lambda, \quad w\ll \Lambda
    \label{umn}
    \end{eqnarray}
The behavior of $U(x)$ at different $I$ and $H$ is shown in Fig. 2.
The transport current tilts $U(x)$, reducing the barrier
maximum and shifting its position $x_0(I)$ toward the film edge.
The barrier disappears at $I=I_s$ for which $x_0(I_s)\sim\xi$.
In turn, the magnetic field at $I=0$ leaves $U(x)$ symmetric, but can produce a
minimum in $U(x)$ at $x=w/2$. There are 3 characteristic fields: $H_b$ at which the
minimum in $U(x)$ appears, the lower critical field $H_{c1}$ at
which $U(w/2)=0$, and $H_s$, at which the edge barrier
disappears. These critical currents and fields can be
calculated from the equation $U^\prime(x_0)=0$.

We calculate $E(I)$ for $T<T_{BKT}$,  $H<H_{c1}(I)$ and $I<I_s(H)$
so that the voltage $V$ results from
thermally-activated hopping of vortices and antivortices over the
barrier $U^\pm(x)=U_0-U_m^\pm$. Here $U_m^-(x)$ for antivortices is given by
Eq. (\ref{umb}) with $H\to -H$ and $\sin^{-1}(x/w)^{1/2}\to
\cos^{-1}(x/w)^{1/2}$ or by Eq. (\ref{umn}) with
$H\to -H$ and $x\to w-x$.
    \begin{figure}                  
    \epsfxsize= 0.8\hsize
    \centerline{
    \vbox{
    \epsffile{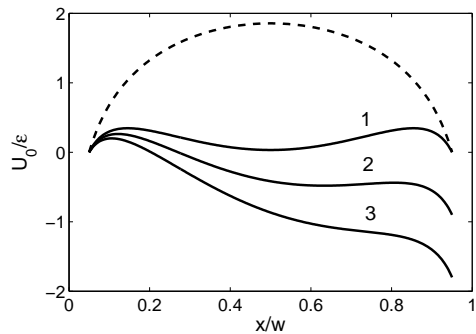}
    }}
    \caption{The vortex energy $U(x)$ given by Eqs. (\ref{u0}) and (\ref{umn}) for a strip with $w\ll\Lambda$
    and $w=20\xi$. The dashed line shows $U_0(x)$, and the solid lines show
    $U(x)$ for $H\phi_0w^2/8\pi\Lambda\epsilon=9$ and different currents, $\phi_0I/c\epsilon$: $0$ (1);
    $1$ (2), and $2$ (3).}
    \label{Fig.2}
    \end{figure}
The mean drift velocities $v_\pm$ of vortices and antivortices
follow from the solution of the Fokker-Planck equation with a
constant probability current~\cite{ah}:
    \begin{eqnarray}
    1=\frac{\eta v_\pm}{wT[F_\pm(w)-F_\pm(0)]}\int_\xi^{w-\xi}dxF_\pm(x)\times \nonumber \\
    \left[\int_0^xdy\frac{F_\pm(0)}{F_\pm(y)}
    +\int_x^wdy\frac{F_\pm(w)}{F_\pm(y)}\right]
    \label{v}
    \end{eqnarray}
where $\beta=\epsilon/T$, $F_\pm(x)=\exp[(U_m^\pm(x)-U_0(x))/T]$, so
that $F_+(0)=F_-(w)=1$, and $F_+(w)=F_-(0)=\exp(\phi_0I/cT)$. The
integral over $x$ is cut off on the scales of the vortex core size,
and the condition $T<T_{BKT}$ implies that $\beta>2$.  If $I\ll
I_s$, where $I_s$ for $w\ll\Lambda$ is of the order of the depairing
current, the x-integral is determined by the vicinity of the edges.
Indeed, for $x\approx 0$, the self-energy $U_0(x)\simeq
\epsilon\ln(x/\xi)$ is dominated by interaction of the vortex with
the nearest image, thus $F(x)\approx (\xi/x)^\beta F(0)$, the first
y-integral in the brackets is negligible and the lower limit of the
second y-integral can be set to $x=0$. Doing the same for $x\approx
w$, we obtain the factor $2F(0)F(w)\xi /(\beta-1)$ after integration
over $x$.

The velocities $v_\pm$ are proportional to the mean electric field
$E\simeq\phi_0(v_+ - v_-)/wc\xi$. This follows from the Joule power
$IV=\phi_0I(v_+ - v_-)L/\xi wc$ produced by the driving force
$I\phi_0/wc$ to move a vortex across the film and
multiplied by the number $\simeq L/\xi$ of
statistically-independent edge sites available for uncorrelated
vortex entries in the strip of length $L$. Using the Bardeen-Stephen
expression for $\eta\simeq d\phi_0^2/2\pi\xi^2c^2\rho_n$
in Eq. (\ref{v}), we obtain \cite{core}
    \begin{eqnarray}
    E=\frac{\pi c\rho_nT(\beta-1)}{d\phi_0}\bigl[1-e^{-I\phi_0/cT}\bigr]\bigl[Z_+^{-1}+Z_-^{-1}\bigr],
    \label{ej} \\
    Z_\pm=\int_0^we^{U^\pm/T}dx. \qquad\qquad
    \label{z}
    \end{eqnarray}
The behavior of $E(I,T,H)$ described by Eq. (\ref{ej}) is shown in
Fig. 3: $E(I)$ is ohmic for $I\phi_0\ll cT$ and nonlinear at higher
$I$. The ohmic $E=R_vI$ at $H=0$ is quantified by the Arrhenius-type
resistance $R_v\propto (-U_a/T)$ per unit length, for which Eqs. (\ref{u0}),
(\ref{ej}) and (\ref{z}) give:
    \begin{equation}
    R_v=\frac{2\pi^{3/2}\beta\rho_n\Gamma(\beta/2)}{dw
    \Gamma[(\beta-1)/2]}\left(\frac{\pi\xi}{w}\right)^\beta,\qquad w\ll\Lambda
    \label{rv}
    \end{equation}
where $\Gamma(x)$ is the gamma-function. The barrier hight, $U_a =
\epsilon \ln(w/\pi\xi)=U_0(w/2)$ depends logarithmically on $w$ in
accordance with Eq. (\ref{u0}).  For $\beta\gg 1$, Eq. (\ref{rv})
yields
$R_v\simeq(\sqrt{2}\rho_n/dw)(\pi\beta)^{3/2}(\pi\xi/w)^\beta$, much
smaller than the normal resistance $R_n=\rho_n/dw$. In wide films
$w\gg\Lambda$, the barrier $U_a \simeq \epsilon \ln(\Lambda/\xi)$
becomes independent of $w$.

For $2\phi_0I> \pi cT$, or $\phi_0Hw >2\pi T$ the change of the
barrier shape $U(x)$ shown in Fig. 2 results in a strongly nonlinear
and field dependent $E(I,H)$, which can be calculated numerically
from Eqs. (\ref{inteq}) and (\ref{ej}) for any ratio $w/\Lambda$,
and analytically for both limits $w\ll \Lambda$ and $w\gg\Lambda$.
For instance, in wide films at $I_s\xi/\sqrt{dw}\ll I\ll I_s$, the
fluxon hopping is limited by the small barriers near the edges:
$U^+(x)\simeq \epsilon\ln(x/\xi) - \phi_0(2I/\pi
c+Hw/2\pi)\sqrt{x/w}$ at $x\ll w$. For $H\gg 2\pi T/w\phi_0$, the
antivortex channel is suppressed, $Z_+\ll Z_-$, so Eqs. (\ref{ej})
and (\ref{z}) yield:
    \begin{equation}
    E=\frac{\pi\rho_ncT(\beta-1)(\xi/w)^\beta}{2dw\phi_0\Gamma(2\beta+2)}\left[\frac{\phi_0}{\pi
    T}\left(\frac{2I}{c}+\frac{Hw}{2}\right)\right]^{2\beta+2}
    \label{ejp}
    \end{equation}
In the limit $I\phi_0\ll cT$, but $H\phi_0 w\gg 2\pi T$, the ohmic
resistance $R_v$ strongly depends on $H$:
    \begin{equation}
    R_v\simeq \frac{\pi\rho_n(\beta-1)}{2dw\Gamma(2\beta+2)}
    \left(\frac{\xi}{w}\right)^\beta\left(\frac{\phi_0Hw}{2\pi
    T}\right)^{2\beta+2}.
    \label{rvp}
    \end{equation}
For $H\ll 4I/c$, but $2I\phi_0\gg \pi c T$, the vortex and
antivortex channels yield the power-law $E(I)$:
    \begin{equation}
    E=\frac{\pi\rho_ncT(\beta-1)}{dw\phi_0\Gamma(2\beta+2)}\left(\frac{\xi}{w}\right)^\beta\left(\frac{2\phi_0
    I}{\pi cT}\right)^{2\beta+2}.
    \label{ejb}
    \end{equation}
For narrow film $w\ll\Lambda$ at $H=0$, the integral in Eq.
(\ref{z}) can be evaluated analytically for all
$I<I_s$:
    \begin{equation}
    \!\!E=\frac{4\pi\rho_ncT(\beta-1)}{d\phi_0w\Gamma(\beta+1)}\left[\frac{2\pi\xi}{w}\right]^\beta\!
    \bigl|\Gamma\bigl(1+\frac{\beta}{2}+i\gamma\bigr)\bigr|^2\!\sinh\pi\gamma
    \label{exact}
    \end{equation}
where $\gamma=\phi_0I/2\pi cT$. In the limit $\gamma\ll 1$, Eq.
(\ref{exact}) reproduces Eq. (\ref{rv}), but for $\gamma\gg\beta/2$,
that is, $J_0\xi/w \ll J< J_0$ where $J_0=
c\phi_0/8\pi^2e\Lambda\xi$ is of the order of the sheet depairing
current density,  Eq. (\ref{exact}) gives
    \begin{equation}
    E=\frac{2\pi\rho_n(\beta-1)}{d\Gamma(\beta+1)}\left(\frac{\phi_0\xi J}{cT}\right)^\beta J.
    \label{uni}
    \end{equation}
This power-law $E(J)$ can also be obtained in the same way as Eq.
(\ref{ejb}) by expanding $U(x)$ near the film edges. Notice that $E(J)$
given by Eq. (\ref{uni}) is independent of $w$ because, once the
vortex overcomes a narrow ($\ll w$) edge barrier shown in Fig. 2,
its subsequent viscous motion across the film is no longer
thermally-activated.

It is instructive to compare Eqs. (\ref{ejb}) and (\ref{uni}) with
the electric field $E_2\sim (\rho_nJ/d)(J/J_0)^{2\beta}$ produced by
the uniform BKT dissociation of vortex-antivortex pairs above the
critical size $\ell_c=2\epsilon c/\phi_0J$~ \cite{hn}. For narrow
films at low temperatures, ($w\ll\Lambda,~\beta\gg 1$), we can use
$\Gamma(z)\simeq (2\pi/z)^{1/2}e^{-z}z^z$ in Eq. (\ref{uni}) and
obtain
    \begin{equation}
    E_2/E\sim (J/2J_0)^{\beta}/e\sqrt{2\pi\beta}
    \label{est}
    \end{equation}
Hence, for $\beta\gg 1$, the virtual vortex-image unbinding
dominates over the uniform pair dissociation except in the region
$T\approx T_{BTK}$ of the genuine BKT behavior. In wide films, the
single-vortex contribution $E/E_2\sim(w/\xi)^{\beta}\gg 1$ is
further enhanced  by the singularities of the Meissner current at
the edges. As an illustration Fig. 3 shows $E(I)$ calculated from
Eq. (\ref{ej}), which gives $E> (10^2-10^3)E_2$ in the region where
$\ell_c<w$. Moreover, $E(I)$ due to the edge vortex-image unbinding
exhibits all characteristic features of the BKT nonlinear transport
in a finite size film: the ohmic $E(J)$ below the critical current
$I_c$ followed by the power-law $E=RJ(J/J_0)^{\alpha_1}$ for
$I>I_c$. Here the exponent $\alpha_1$ varies from
$\alpha_1=2\beta+1$ for wide films to $\alpha_1=\beta$ for narrow
films, while the uniform pair unbinding gives $\alpha_2=2\beta$
\cite{hn}. The similarity of $\alpha_1$ and $\alpha_2$ in wide films
results from the edge Meissner singularity of $J(x)$, which
increases $\alpha_1$ as compared to $\alpha_1=\beta$ for a uniform
$J$. The critical current $I_c$ is estimated from the condition that
the maximum of $U(x)$ at $x=x_0(I)$ shifts from the film center at
$I\ll I_c$ to the edge at $x_0\ll w$ for $I\gg I_c$. For a narrow
film, $I_c$ defined by $x_0(I_c)=w/4$ in Eqs. (\ref{u0}) and
(\ref{umn}) is:
    \begin{equation}
    I_c(H)= \frac{c\phi_0}{16\pi\Lambda}\left(1-\frac{H}{H_0}\right),\qquad
    H_0=\frac{\phi_0}{w^2},
    \label{ic}
    \end{equation}
so that $I_c(0)$ is independent of $w$, but both $J_c(0)=I_c/w\sim
J_0\xi/w$ and $H_0$ increase as $w$ decreases. The same $I_c(0)=\pi
c\epsilon/\phi_0$ is obtained, defining the nonlinearity onset from
the condition $l_c=2w/\pi$ equivalent to $\gamma=\beta/2$ in the
argument of the gamma-function in Eq. (\ref{exact}).

    \begin{figure}                  
    \epsfxsize= 0.95\hsize
    \centerline{
    \vbox{
    \epsffile{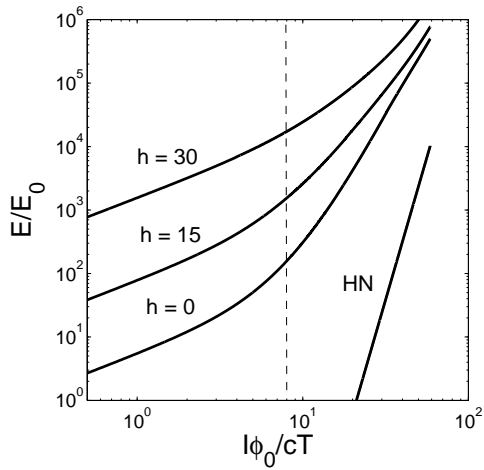}
    }}
    \caption{$E-J$ curves calculated from Eq. (\ref{ej}) for $\beta=4$, $w=20\xi\ll\Lambda$ and
    different fields $h=H\phi_0w^2/8\pi\Lambda T$.
    Here $j=I\phi_0/cT <j_s=2\beta w/e\xi$, and $E_0=\pi c\rho_n T(\beta-1)(\pi\xi/w)^\beta/dw\phi_0$.
    The line labeled by HN shows the Halperin-Nelson result, $E_2(I)$~\cite{hn}. The critical
    pair length $\ell_c$ exceeds $w$ in the region $j < 2\beta$ left of the dashed line.   }
    \label{Fig.3}
    \end{figure}

The results presented above indicate that $E(J)$ can be tuned by changing
the film geometry. For instance, if a uniform $J(x)$ is produced
in a wide film, the exponent $\alpha_1=2\beta+1$ would decrease to
$\alpha_1=\beta$. This could be implemented by using
ferromagnetic/superconducting structures \cite{vgk,sf}, in which a
thin film strip is placed perpendicular to ferromagnetic screens to
eliminate the singularity in $J(x)$ \cite{sf}. Another possibility
is to use a thin film tube in a parallel field, which produces
uniform azimuthal screening currents $J=cHd/4\pi\lambda$ driving
vortices along the tube. Because of the negligible demagnetization
factor of this geometry, $J(x)$ for large tubes of length
$L\gg\Lambda$ and diameter $D\gg\Lambda$ does not contain the
Meissner edge singularities characteristic of wide films in a
perpendicular field. Such a tube would have a mixed $E(J)$ controlled by
$U_0(x)$ of a wide strip, but a uniform current drive like in a narrow film.

Film and ring structures make it possible to probe $E(J)$ by
magnetic relaxation measurements well below the nV voltage
sensitivity \cite{lobb} of transport measurements. In this case
$H(t)$ is ramped up and then stopped, after which the magnetic
moment $M(t)=I(t)D/2c$ is measured. For $\pounds I\gg \phi_0c$,
relaxation of $I(t)$ in a ring or a tube is described by the circuit
equation $\pounds\dot{I}=-\pi c^2D RI(I/I_0)^\alpha$, where $\pounds$
is the self-inductance. The solution of this equation,
$I(t)=(\tau/t)^\alpha I_0$  with $\tau=\pounds/\pi c^2DR\alpha$,
enables extracting $\alpha(T)$ from flux creep measurements after
some initial transient time \cite{creep}.

In conclusion, thermally-activated fluxon hopping mediated by
unbinding of single vortices from their edge antivortex images can
mimic the nonlinear resistive behavior of a uniform BKT state. Our
results predict a strong dependence of $E(J,H,T)$ on temperature,
magnetic field and the sample size. This offers a possibility of
tuning the behavior of $E(J)$ by changing the film geometry or by
incorporating magnetic structures.

We thank B. Rosenstein for hospitality at the International Center
of Theoretical Sciences at the Hsing-Hua University, Taiwan, where
this work was started. The work was also supported by the NSF grant
DMR-0084173 with support from the state of Florida (AG) and by the
U.S. Department of Energy Office of Science through contract No.
DE-AC02-06CH11357 (VV).  We are grateful to B. Altshuler and T. Baturina for useful
discussions.


\end{document}